\documentclass[12pt]{article}
\usepackage{amssymb,amsmath}
\usepackage[authoryear]{natbib}
\usepackage{amsbsy,amsfonts,enumerate,epsfig,graphicx,natbib,rotating,subfigure}

\begin{document}

\markboth{M.A.Virasoro}{Time evolution of the Intraday non-Gaussianity}

\title{\textit{Non-Gaussianity of the Intraday Returns Distribution}: its evolution in time}

\author{M. A. Virasoro\\
Instituto de Ciencias, Universidad Nacional de General Sarmiento.\\
Dipartimento di Fisica, Universit\`a di Roma ``La Sapienza''.
\\
Email: miguel.virasoro@roma1.infn.it}
\maketitle

\begin{abstract}
We find a remarkable time persistence of various proxies for the kurtosis (p-kurtosis) of the intraday returns distribution for the S\&P500 index and this permits a significant measure of their evolution from 1983 to 2004. There appears a long time scale dramatic variation of the p-kurtosis uncorrelated with the variation of the volatility thus falsifying any hypothesis of a universal shape for the probability distribution of the returns.  A large increase in the kurtosis anticipates the October 87 crash. During the years 1991-2003 it continuously decreases even when the volatility grows during the dot-com bubble. We propose some speculative interpretations of these results.
\end{abstract}

\section{Introduction}

The availability of high frequency intraday data generated by Stock Markets presents a challenge: how to manipulate that amount of data to derive the maximum amount of meaningful results from it. Economists (econometricians), psychologists and even physicists, with their respective methods and working hypotheses, are confronting this challenge. This paper is still another try.  We will show that there is another collective variable that characterises the state of the market in addition to \emph{index quotes} and  \textsl{volatilities} and that the time evolution of this variable can be measured in the sense that its signal is larger than the noise. In this paper, for concreteness, \emph{state of the market} refers to its principal component as seen through the observation of the S\&P500 index. 

The alert reader may object what is the purpose of measuring something if it is not first embedded in a theoretical framework. We postpone this question because our motivation originated in a long detour through epistemological/psychological considerations and is highly speculative. Still here we can say the following: the market agent takes decisions in the middle of uncertainties. Since the work of  \citet{knight21} we believe there are 2 types of uncertainties: \emph{risk} that can be modeled through consistent probability distributions, and proper \emph{uncertainty} or \emph{ambiguity} that is more resistant to formalization. Still \citet{bewley02}, \citet{gilboa09}, \citet{schmeidler89} have tried to formalize this kind of uncertainty and derived from it some idiosyncratic behaviors: inertia, indeterminateness, uncertainty aversion. It has also been proposed that there should be traces of it in the market behavior (\citet{basili01}, \citet{mukerjit03} and references therein). This context then justifies, at least partially, the search for a new parameter characterizing the market state because just one, namely the value of volatility at time $t$ could not measure both risk and uncertainty at time $t$ \footnote{We do not rule out that one could build this parameter exploiting, for instance, the time derivative of the volatility. See discussions below}  

The following general framework is adopted: the data of the S\&P500 returns time series are understood as extracted from a statistical sample obeying a distribution whose parameters vary with time. This is not controversial and in fact the time dependences of two of these parameters, the volatility and the mean return, are usually discussed in the literature. We will add, at least, a new one. The approach is otherwise non-parametric.

In order to organize the data one defines two time scales. The first one $\delta t$ is the interval of time that defines the returns: the return at time $t$, $r_t=\log(p_{t+\delta t})-\log(p_t)$. The second one is the temporal window $\Delta t$ that bunches together the data from the time series to build a (time dependent) statistical sample. For time scales below such value the  variations of the distribution parameters is, by definition,  not observable. For time scales larger their measurability is not given but depends, of course, on the  signal to noise plus systematic error ratio.

In the great majority of studies undertaken on these type of time series the program is to fit a specific dynamic model with the final goal of partially forecasting the future evolution. The family of ARCH models have been extensively used with this purpose (see for instance the review: \citet{bollerslev94}. In this approach parsimony is essential. If the model has too many parameters, one falls into the overfitting trap and the extrapolation of the model gives wrong forecasts. On the other hand, the choice of a model just because of its simplicity, flexibility and/or parsimony does not justify leaving aside a necessary preliminary and open-minded exploration of the data.

With this observation in mind we here take a step backwards and try simply to find parameters that enable an interesting narration of the history of the time series. We postpone on purpose any pretension to a dynamical description of the data.

But apart from this difference in strategy, what we will propose can be seen as a straightforward generalization of the \citet{anderetal03} proposal of using realized volatilities. In few words we will take into consideration something like a realized kurtosis, that is the kurtosis of the intraday returns distribution. More precisely a proxy (or proxies) that similarly measure the non-Gaussianity of the intraday return distribution but are less prone to the small sample biases and/or easier to measure\footnote{We will precede words with a p to refer to a chosen proxy i.e. \emph{p-kurtosis} , \emph{p-volatiility}. Furthermore in all cases we normalize these p-kurtosis by substracting the corresponding value taken by a pure Gaussian}. \citet{bouchpottbk} have thoroughly discussed how the kurtosis of the returns distribution decreases slowly with $\Delta t$ as a consequence of heteroskedasticity. Once the time interval (and thus the finite sample) is determined the p-kurtosis to be measured could partially come from an intra-$\Delta t$ variation of the volatility. We will not try to disentangle these different contributions but we will still correct for the systematic, periodic, volatility intraday variation connected with the opening and closing hours of the market \citep{andersenb97}. We will show that our p-kurtosis can be measured, is time dependent, is persistent and to a large extent uncorrelated with the p-volatility.

There have been many studies done in the framework of ARCH models and its relatives that include or allow for time dependent conditional kurtosis (see in particular \citet{hansen94,harveys99,jondeaur03,brooksetal05,dark10}). The conditional kurtosis in ARCH-like models refer to the daily returns and not to the intraday shorter period returns. In addition its identification strictly depends on the parametric model because these models also include day to day volatility variation and disentangling both effects is model dependent. There is no necessary connection between this conditional kurtosis and our realized p-kurtosis. In any case we remark that those studies show that the null hypothesis that the conditional kurtosis is constant in time is rejected but do not add information about the actual time dependence including the possible time persistence.

In the idealized world of continuous time models the attention necessarily focuses on  Jumps  plus Diffusion models\citep{merton76}. With the advent of high frequency data it becomes possible to calibrate these models with a successful separation of the jumps from the Brownian continuous component(\citet{mancini01}, \citet{barndorffs04}). What distinguishes a jump from a continuous contribution is the sudden variation of the return. For instance in one line of research(Mancini (2001), (2004) and (2009)) if the price variation in a small time interval $h$ is larger than a specified deterministic \emph{threshold} function of $h$ (for instance $h^\beta$ with $\beta>1$) then that variation is attributed to the jump component. If $h$ could be taken arbitrarily small then these methods would allow to identify all jumps. This is obviously impossible because when $h= \delta t$ is chosen too small the real market process has nothing to do with a continuous time stochastic process (see \citet{lilloetal} for a discussion of some of the complexities of price formation at micro time scales). Therefore only jumps large enough can be identified \citep{andersenetal07} and these in general are rather rare and do not show persistence\footnote{See however \citep{mancinir11}, specifically Figure 5.4 where clustering is apparent (C.Mancini: private communication).}.  On the other hand the non-Gaussianity of the intraday returns that interests us (not the one of the daily returns) would originate in these models both from the small and large jumps and also from the intraday time variation of the volatility.  Our approach is different because we do not disentangle these factors and in compensation discover an observable that has striking persistence in time and therefore allows for a better filtering of the noise and can be interpreted as a market state collective variable. There are, in this last respect, similarities with the contemporary work of Bollerslev and Todorov (2011a and b): these authors do start discussing the same specific time continuous model but they use it instrumentally to propose a non-parametric way of studying the data. Their reliance on the medium jumps to estimate the large ones and our use of proxies of the kurtosis are similar and their interpretation of it as a signal of fear also implies that they consider it as a market state colective variable. 

We want finally to mention similar attempts to define new \emph{market state variables} in the larger context provided by cross-sectional distributions of share returns. In particular \citet{borland} points to some interesting signals that characterize the advent of a crisis. Due to a self organization that increases the correlation among shares  cross-sectional kurtosis decreases and the dispersion increases in times of panic while the opposite occurs during normal times. In a related paper \citet{bouchauda} confirm these results and in addition find new regularities affecting in particular the seasonal intraday variability of returns that suggest a more general anti correlation between kurtosis and dispersion.

In this paper we will analyze the 5 minutes returns of the S\&P500 index from 1 February 1983  to the 30th November 2004 for a total of approximately 5472 days. We will show that the evolution of the kurtosis proxy tells an interesting story that is not contained in the daily quotes and volatilities. In particular we will discuss the first 1500 days and specifically the approach to the Black Monday Crash (19-October-1987). We choose $\Delta t$ equal to one day. Therefore we will have 72 quotations up to September 1985 that then increases to 78. 

The paper is organized as follows: in the next section we present the data and define the observables that we try to measure. We also discuss our reasons to choose the 1 day period and the need to eliminate the first and last hours of each session to derive cleaner data. In Section 3 we present the results on the time autocorrelation of the p-kurtosis and thus justify taking a 60 days exponential moving average and  in a subsection we discuss and show the historical evolution of the p-kurtosis. In section 4 we present the Montecarlo simulations used to estimate errors in our measurements. We also present 2 alternative definitions of the p-kurtosis, one based on quantiles following \citet{moors88} and another one, somehow noisier but sensitive only to the excess number of returns on the tails of the distribution. Finally the last section is dedicated to the conclusions and discussions.
  
\section{The Data}

As stated in the introduction our data consists of the 5 minutes S\&P500 index quotations from February 1983 to the 30th November 2004 for a total of approximately 5472 days. 

\textit{Notation:} If $q_{i,j}$ are the index quotation on day $i$, period $j$,  the discretely sampled returns are denoted: 
\begin{equation}
r_{i,j}=\log(q_{i,j+1})-\log(q_{i,j})
\end{equation}

As discussed in the next paragraph we will disregard the data corresponding to the first and last hours of every market session. Furthermore, we are interested in the time evolution of the degree of non-Gaussianity of the intraday distribution and therefore we normalize the returns dividing by the average value of the absolute value return on each day (similar to the so called {\it standardized} \citep{andersenetal00} returns).
\begin{equation*}
\widehat{r}_{i,j}=\frac{r_{i,j}}{(\sum_{m=1,np_i}{|r_{i,m}|})/np_i}
\end{equation*}
where $np_i$ is the number of 5 minute periods on day $i$ minus the 24 periods corresponding to the opening and closing hours. We will use the notation $\overline{(a_{i,j})_j}$, $\overline{(a_{i,j})_{i=i_1,i_2}}$ to indicate averages over periods on a fixed day and averages over a specific days interval respectively.

We concentrate our attention on the following parameters of the intraday distribution:

\begin{eqnarray*}
V_i=\overline{(|r_{i,j})|)_j}\\
K_i=\overline{(\widehat{r}_{i,j}^{\;2})_j}-\frac{\pi}{2}
\end{eqnarray*}
These parameters \emph{p-volatility} and \emph{p-kurtosis} are proxies for the usual ones. The p-kurtosis is zero for a normal distribution and grows if the latter becomes more leptokurtotic.

We have  followed \citet{andersenb97} in the analysis of the variation of the volatility near the opening and closing hours. For this specific calculation we eliminated those days with abnormally short sessions, in total 13 days. Even then we had days with 71, 72, 77 and 78 periods. Assuming that it is the distance to the opening and closing moments that is relevant we uniformized the data simply selecting  71 periods dropping those near midday. For each one of these 71 daily periods we calculated the average of the absolute return over all days to check and confirm that in the first and last hours there is a large increase in the volatility. This variation will contribute to the calculation of the p-kurtosis and obscure possible more interesting contributions that reflect an evolving market state. Therefore we eliminated from the data the first and last hour in every session. We then took partial averages of the remaining hours over bunches of 500 days  as well as a full 5000 days average to check whether there could be some important variation with time. In Figure 1 we show the plots of the 5000 days average (Fig. 1a) and 4 partial averages corresponding to the periods 1-500, 1000-1500, 2500-3000, 4000-4500 of the following quantities:
\begin{equation*}
\overline{(|r_{i,j}|)_{i=1+(k-1)500,k 500}}\qquad \textrm{with}\;\; 1\le j\le 47\; \textrm{;}\;k=1, 3, 6, 9
\end{equation*}
\begin{figure} 
\begin{center} 
\subfigure[]{ 
\resizebox*{5cm}{!}{\includegraphics{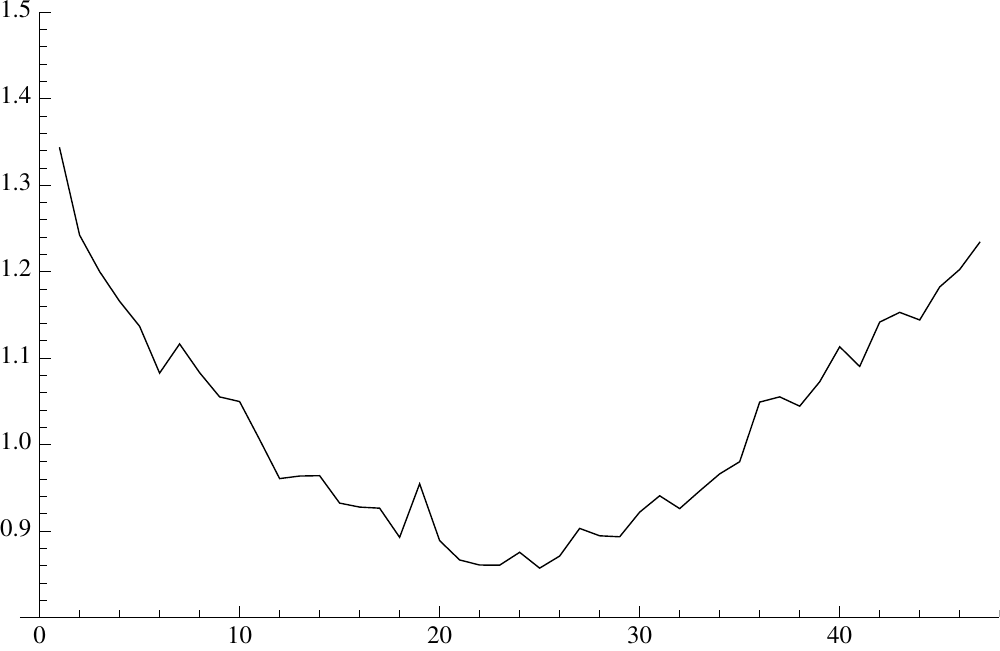}}}%
\subfigure[]{ 
\resizebox*{5cm}{!}{\includegraphics{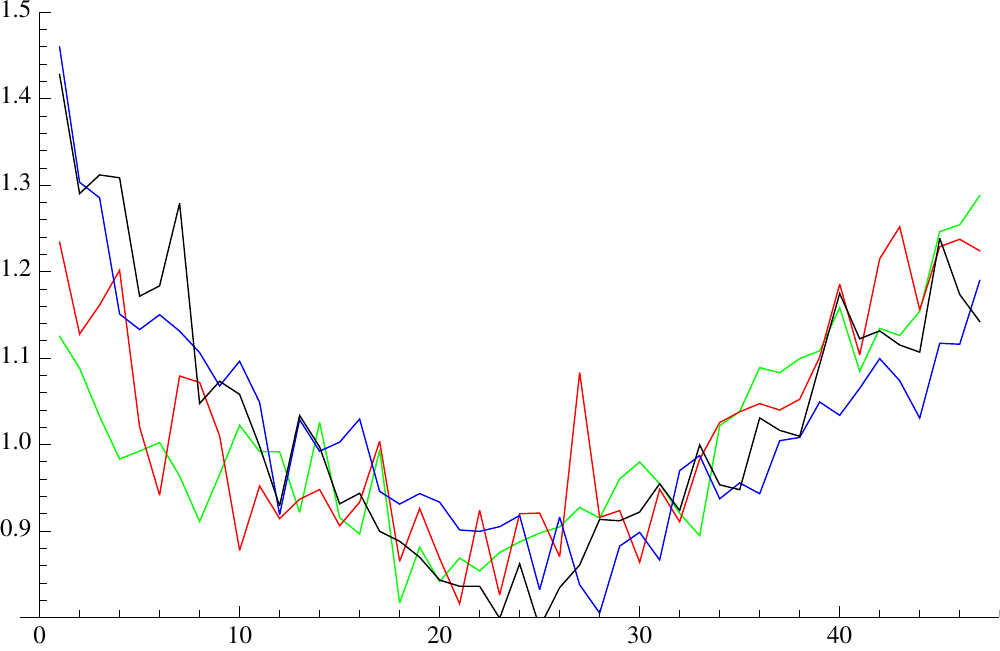}}}%
\label{sample-figure}
\end{center} 
\caption{Fig 1a shows a 5000 average of the normalized intraday volatility once the first and last hours have been cancelled. Fig 2a shows the averages over the periods: 1-500 (green), 1000-1500 (red), 2500-3000 (blue), 4000-4500 (black).}
\end{figure}
The profiles are rather similar, the time dependence is insignificant. There remains a residual intraday variation of the order of 30\%. The estimated value of the p-kurtosis thus generated is $K_{Fig.1}=0.027$ so values below or of this order are not significant.

\subsection{Why the choice $\Delta t=1\, day$}
With the cuts proposed in the last paragraph we are left with a modest number of time intervals each day. Therefore errors on the estimates of allegedly ``true" (whatever that means) PDF's parameters could be rather large. An immediate way to evade this would be to take $\Delta t$ of the order of a week or a month. Then variation of the daily volatility inside that time interval will generate additional p-kurtosis. There is therefore a trade-off: kurtosis in those scales is not that interesting because there are many mechanisms illustrated in as many models that explain variation of volatility in such longer periods. For instance the herd effect/contagion \citep{luxm00}, learning by agents \citep{brianetal96}, rational expectation of a possible, unrealized \emph{Peso Problem} \citep{veronesi03} all produce both clustering, persistence and variation of the volatility in time scales larger than a day. But unless we stretch unrealistically those mechanisms and models, they -and to the best of our knowledge all other models- do not explain the non-gaussianity at the intraday frequency. 

There remains still the other \textit{Null} hypothesis i.e. that these fluctuations are due to exogenous news arriving to the market. This possibility has been carefully considered and ruled out by the work of  Joulin \textit{et al} (2009) who cross-examined news arrivals and the observed jumps. In section 4 we will show that our measurements are on the bulk compatible with theirs and add some new evidence to their conclusions.

\section{Persistence and evolution of the p-kurtosis parameter}
\subsection{Persistence}
If we calculate the time autocorrelations of the p-kurtosis and the p-volatility \underline{as if} the data were stationary: 
\begin{eqnarray*}
C^{(K)}_t=\frac{\overline{(K_i K_{i+t})_{i=1,ntd-t}}}{\overline{(K_i K_i)_{i=1,ntd}}}  \\
C^{(V)}_t=\frac{\overline{(V_i V_{i+t})_{i=1,ntd-t}}}{\overline{(V_i V_i)_{i=1,ntd}}}
\end{eqnarray*}
we obtain the plots shown in Fig. 2: the correlation of the p-kurtosis falls to 0.3 after just one day but later shows remarkable persistence similar to the persistence demonstrated by the p-volatility.  It is reasonable to assume that the big fall is to some extent due to errors introduced by our manipulation of the data, a fact that will be checked with Montecarlo simulations in Section 4. At the same time the persistence suggests averaging the p-kurtosis over many days to reduce this noise.
\begin{figure}
\begin{center}
\includegraphics{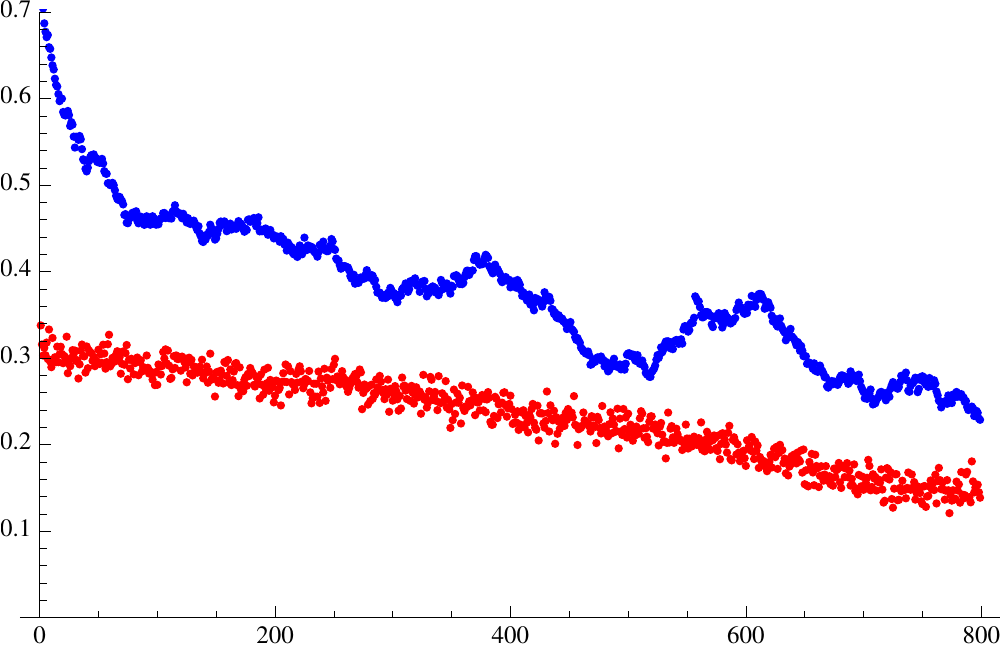}
\end{center} 
\caption{Autocorrelations of the p-Kurtosis (red) and the p-Volatility (blue).}
\end{figure}

A natural question arises at this point: are the p-kurtosis and the p-volatility strongly correlated. If the answer is positive then there would be some hope of finding a universal profile for the probability distribution of the returns. If they are not, the two parameters p-Volatility and p-Kurtosis must have different interpretations. We have calculated these correlations in different 600 days periods and it is in general small and varies wildly, being sometimes positive and sometimes negative. This will be clarified in the next subsection.

\subsection{Historical profile of the p-kurtosis and the p-volatility}

As the apparent scale invariant decrease of the correlations indicates, there are multiple time scales relevant in the evolution.

Let us choose averaging both variables over 60 days period (more or less corresponding to a quarter). We use an exponential moving average for this purpose, therefore attenuating both noisy and real variations below that time scale:
\begin{equation}
f_{ma}(t)=\frac{\sum_{i=t-60}^t {e^{\frac{i-t}{60}} f(i)}}{\sum_{i=t-60}^t {e^{\frac{i-t}{60}}}}
\end{equation}

In Fig. 3 we show the moving average of the p-Kurtosis together with the corresponding one of the p-Volatility conveniently normalized so that it fits in the same graph. We can clearly see the persistence of both quantities. At the same time it becomes apparent that there is \underline{no} correlation between the 2 observables at very long time scales: there are periods where both increase or decrease and there are periods when one increases while the other decreases. At shorter time scales there appears to be some interaction between the two quantities. The fluctuations that we see in the figure could have been introduced by our manipulation of the data. In the next Section we will present Montecarlo simulations of a constant probability distribution of returns manipulated exactly as the data. The fluctuating result is shown in Figure 3 (Green curve). The discussion is left to the next Section.

\begin{figure}
\begin{center}
\includegraphics{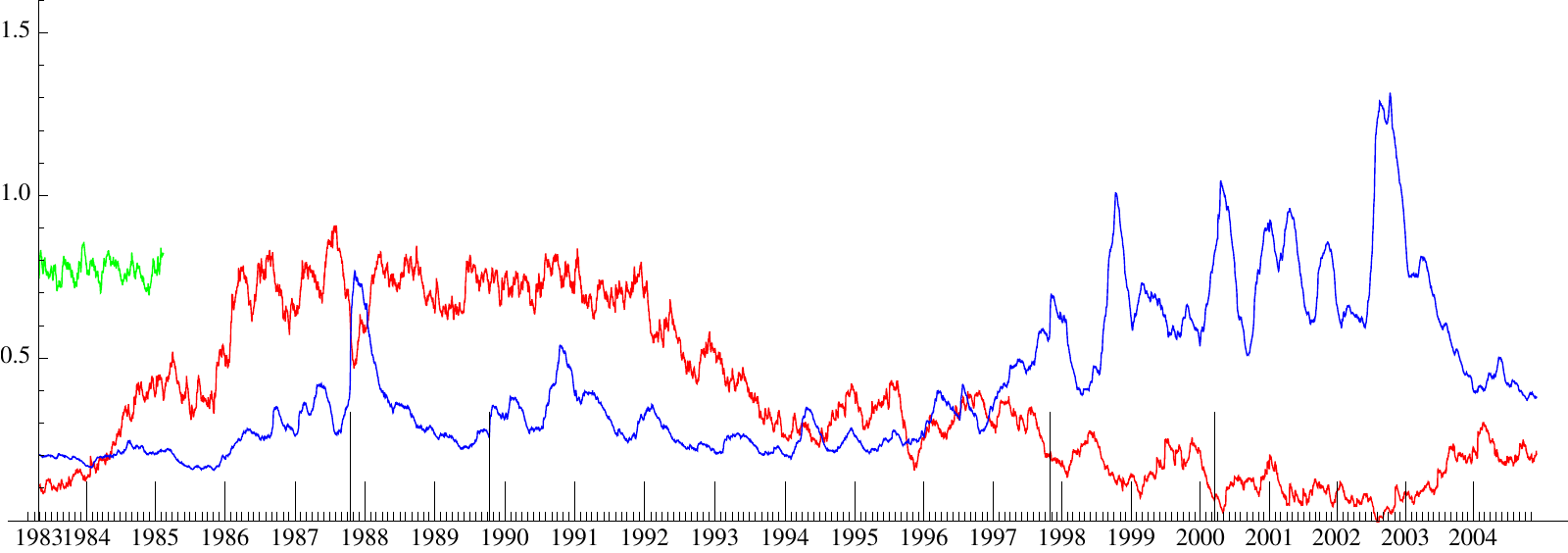}
\end{center} 
\caption{Graphs of the exponential 60 days moving average of the p-Kurtosis (red) and the p-Volatility (blue), this one multiplied by 1000. The green curve represents Montecarlo simulations discussed in Sect.4 The large ticks on the time axis correspond to special dates: 19-10-1987 crash, 13-10-1989 mini-crash, 07-1997 Asian financial crisis, 03-2000 dot-com bust)}
\end{figure}

The unavoidable immediate conclusion of this graph is that the importance of the historical context and narration can not be evaded. If the correlation between p-Volatility and p-kurtosis in periods of 6 months comes out sometimes positive and sometimes negative, the insistence in finding a universal profile for the returns distribution simply crashes against reality.

The second observation is the striking difference in the approach to two different crisis. There is a remarkable increase of the p-Kurtosis that seems to anticipate the October 1987 Crash while there is a decrease of the p-Kurtosis during the approach to the dot-com bubble burst. We will discuss in the last Section the possible reasons for this difference.

\section{Analysis of errors and alternative proxies}

Figure 2 proves beyond any reasonable doubt that the time evolution of the p-kurtosis is real. We do not (and could not) claim that the value measured any day reflects some kind of \emph{``real"} parameter of the statistic ensemble of that day data because our numerical manipulation surely biases the results. In particular the relatively small sample and the normalisation of the data by the daily p-volatility are guaranteed to reduce any kurtosis measure (in the limit of a 1 period sample the measured p-kurtosis will be one; furthermore the contribution of any period to the kurtosis is bounded by the number of periods in the day), Finally the figure shows clear dispersion of the data around a moving average. We want to know whether such fluctuations are real fluctuations of the p-kurtosis or simply pure statistical fluctuations due to the small sample or perhaps a mixture of both.

With those questions in mind we have done some Montecarlo simulations. We assumed that the data could be reproduced by mixing two Gaussians and fitted the 2 parameters: the ratio between the dispersions and the relative amplitude using the p-kurtosis at the maximum and a second parameter:
\begin{equation}
K^{(0)}_i=\overline{(\widehat{r}_{i,j}^{\;1/2})_j}-\frac{\Gamma(3/4)}{\pi^{1/4}}
\end{equation}
where the constant is there so that this parameter be zero for a pure Gaussian distribution. For a leptokurtotic one it is generically negative

In addition we also considered an alternative normalisation of the returns\footnote{we thenk J-Ph. Bouchaud for suggesting this variant}:
\begin{equation}
\widetilde{r}_{i,j}=\frac{r_{i,j}}{\sum_{m,m\neq j}{|r_{i,m}|}/(np_i-1)}
\end{equation}
 
which has the apparent property of biasing in the opposite direction (for samples with just one element the kurtosis becomes infinite). From the data around the maximum value, few weeks before Black October Monday we measure: $K=0.77, K_b=1.16, K^{(0)}=-0.042, K^{(0)}_b=-0.029$. The parameters of the mixed Gaussian were then adjusted to be:
\begin{eqnarray}
a&=&0.80 \quad     \textrm{amplitude of the } N(0,\sigma_1)     \nonumber \\
\sigma_1&=&0.62    \\
\sigma_2&=&2.54    \nonumber
\end{eqnarray}

We generated 5000 samples of 50 periods with this constant in time probability distribution and analysed them with the same method applied to the S\&P500 data. The first 350 points are drawn in Fig.3 (see caption) , The mean values for the simulated $K$'s are: $K^{sim}=0.76, \: K^{sim}_b=1.19, \: K^{(0)\,sim}=-0.042, \: K^{(0)\,sim}_b=-0.027$. The variances of these quantities are:
\begin{eqnarray}
\sqrt{var(K^{sim})}=0.051  \nonumber  \\
 \sqrt{var(K_b^{sim})}=0.078\\  \nonumber
 \sqrt{var(K^{(0)\,sim})}=0.0029\\  \nonumber
 \sqrt{var(K_b^{(0)\,sim})}=0.0027    \nonumber
 \end{eqnarray}
 
 The first of these values can be used to draw error bars in Fig 3 though we find the direct comparison of the simulated curve with the data more illuminating.  We also notice that $K$ and $K_b$ small sample biases are in opposite directions as argued (the $K$ calculated analytically for the mixed Gaussian is 0.91). Finally if we want to believe in extrapolation, the kurtosis derived from the mixed Gaussian distribution is 6.97, consistent with figures appearing in the literature though here, if anything, it corresponds to the maximum value  of a quantity that evolves in time.
 
 Comparing the fluctuations in Fig 3 of the real data and the simulations we can conclude that a significant fraction of them are statistical errors consequence of the manipulation of the data. But still some structure is apparent. For instance some anticorrelation or correlation between bursts of p-volatility and p-kurtosis is visible at least at time scales of the order of several months. The next (future) challenge is to find a way to make this structure significant.
 
 Finally we have experimented with alternative measures of the non-Gaussianity depicted in Fig 4 and 5. Specifically we looked at the \citet{moors88} kurtosis defined using the octiles and a measure that is sensitive exclusively to the tails. For the latter we followed \citet{joulinetal09} and proceeded as follows:
 \begin{enumerate}
 \item On each day we calculated the median and separated the data according to whether they were larger or smaller than the median.
 \item We computed the average value of the return $V_i^{(L)}$ and $V_i^{(R)}$ for both groups.
 \item{We finally counted how many returns fell in the intervals $(-\infty, 4 V_i^{(L)})$ and $(4 V_i^{(R)}, \infty)$}
  \end{enumerate}
  
 \begin{figure}
\begin{center}
\includegraphics{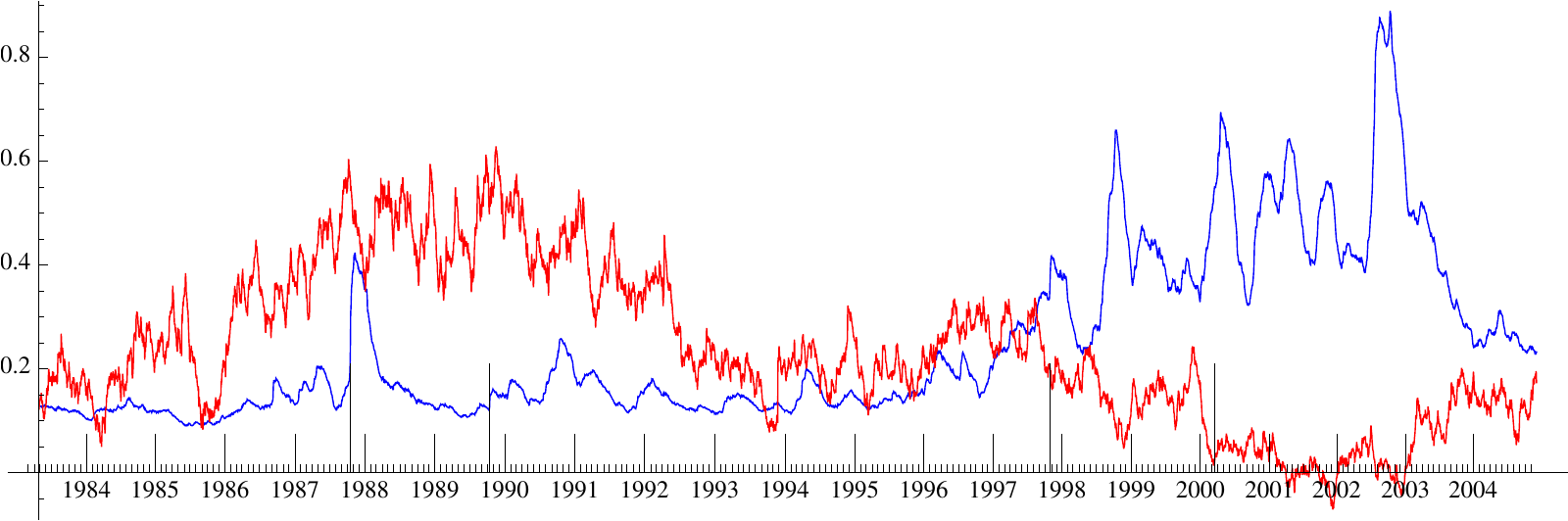}
\end{center} 
\caption{Graphs of the exponential 60 days moving average of the Moors kurtosis defined by the Octiles $O_{i}$ as $\frac{O_7-O_5+O_3-O_1}{O_6-O_2}$. The ticks are as in Fig 3}
\end{figure}
\begin{figure}
\begin{center}
\includegraphics{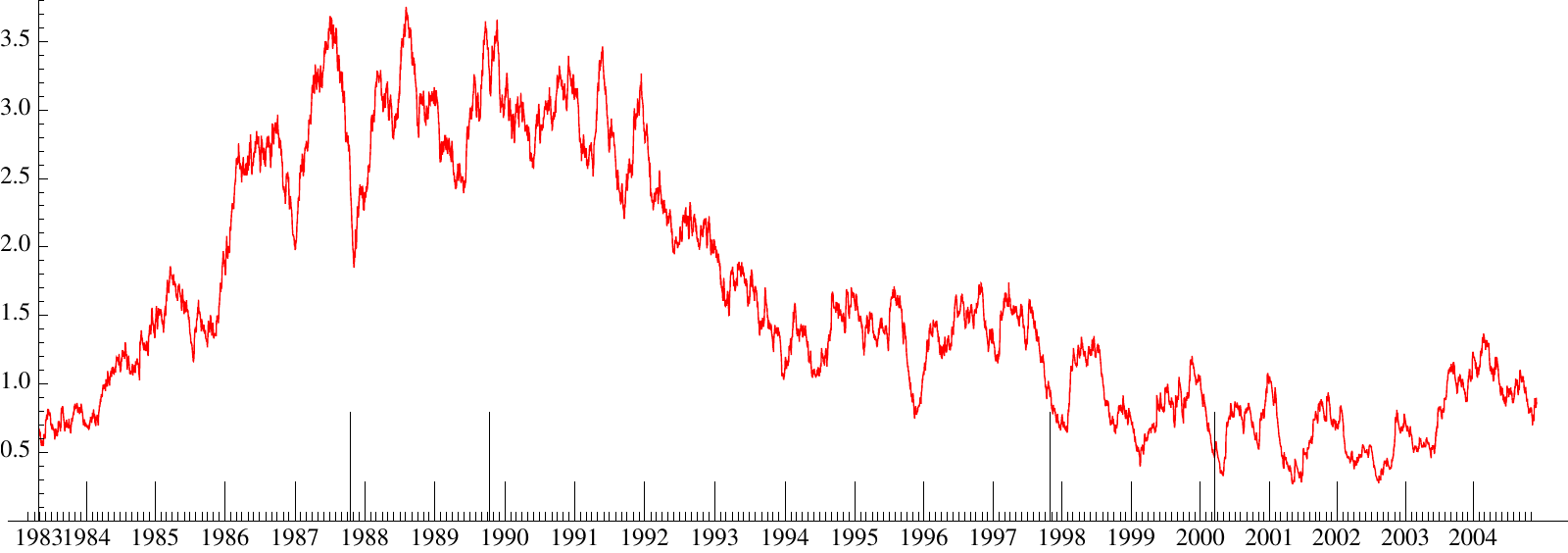}
\end{center} 
\caption{Exponential 60 days moving average of the number of events falling in the tails defined as in the text }
\end{figure}

 This last measure is interesting because it depends exclusively on the tail, does not mix a possible skewness with the kurtotic property and has robust statistical properties. In a pure Gaussian the probability of an event falling in that interval is 0.0014 equivalent to  one event every 14 days. Instead the figure shows that around the maximum, on the year 1987, there were 3 to 4 events per day. This value is compatible with the findings of \citet{joulinetal09} even if it is smaller because it is known that the S\&P500 index is less kurtotic than individual shares. Our analysis adds a new strong argument in favour of that paper's thesis:  news are not necessarily at the origin of the jumps. In fact the months before October 1987 is according to most contemporary witnesses a period of minimum news and still is a period of maximum number of jumps. 
 
 \section{Conclusions and Discussions}
 As told in the introduction the motivation for this work developed in a tortuous path through psychology and epistemology. I sketch here the line of reasoning that will be expanded in a future work.  
 
 At its origin there was a shapeless idea that uncertainty was not fully considered in Economics even if the work of Knight, at least in the interpretation implicit in the formalizations proposed by his followers, was taken into account. This was compounded with the realization that in a larger context (i.e. Agent Based Models and bounded rationality as proposed by \citet{sargent93})  \emph{learning} was identified with \emph{induction} when instead we know that, in order to acquire knowledge (at least \textit{ampliative} knowledge) something else, for instance what C. S. Pierce called  \emph{abduction}, has to be invoked\footnote{"Abduction is the process of forming an explanatory hypothesis. It is the only  logical operation that introduces any new idea; for induction does nothing but  determine a value, and deduction merely evolves the necessary consequences of a  pure hypothesis" \citet{peirce}. See also \citet{hintikka98}} . This mode of inference is essential in many cognitive processes, specifically, in guessing a \emph{frame}, that is which and how many variables are to be considered relevant to a particular phenomena. It is inside that frame where \textit{induction} can work to adjust the model. The infinite variables left outside add noise and the inductive optimum will be reached when that noise appears to be perfectly random. We are left with a useful statistical model but uncertain because the noise could be correlated with a variable outside the frame.   
 
 This uncertainty is \textit{irreducible} inside the frame but is tolerated because of the cost of exploring a very large number of variables (the \textit{curse of dimensionality}). In most contexts the agent totally and rationally ignores it. The attitude however has to change in a competitive environment. Another agent may exploit correlations with variables assumed irrelevant and this will affect non-linearly the model predictions: our agent has to remain aware of the limitations of his/her frame.  We then expect under-reaction of the agent in front of small discrepancies because the trust in the model is limited and over-reaction, i.e. jerkily following the market and restricting the supply of liquidity if the discrepancy becomes large enough. 
 
 We recognize that this is somehow speculative. We would need both an Agent Based Model to check that the effect does not disappear in the bulk and perhaps even classroom experiments to measure the reactions.  
 
 On the other hand the validity of the work exposed in the rest of this paper does not depend on these considerations/motivations. The key result is the surprising long time scale evolution of the different measures of the non-Gaussianity of the intraday data.  As far as we know the observation of the slow increase culminating in the Black Monday October 87 crash as well as the still slower decrease that traverses the whole burst of the dot-com bubble have not been noticed before and deserve some explanation and further study. 
 
 We mention \textit{en passant} that they are compatible with our previous speculations. The dot-com burst was highly anticipated. Private conversations with the protagonists  confirm that they knew they were living through a bubble but the incentives to stay in the market were important. The uncertainty was about the timing of the burst, how to maximize profits before that date and strategies to leave the market when the moment arrived. Those with different predictions were confident in their models and therefore perfectly happy to bet on the outcome in spite of the large risk. They were closely following signals coming from the economy and this produced turbulence but on a longer time scale. Just the opposite was happening before October 1987: at that moment the post-mortem analysis done by  \citet{shiller89} through interviews of the protagonists found that most investors did not pinpoint one specific arriving news as causing the existing recognized state of anxiety. They were worried by the psychology of fellow investors and had a gut feeling that the market was overvalued. The continuous growth of the index since the end of Volcker tightening was not totally understood. In a few words they had diminished trust on their models but did not know how to correct them. In fact we know a-posteriori that the models did not need any correction. 
 
In addition to a more detailed discussion of our motivations many interesting problems remain opened. First of all the connection with the Jumps analysis of Bollerslev and Todorov (2011a and b) that will presumably require repeating our analysis with the S\&P 500 future index which is more commonly used in high frequency data.
 Then also the shorter time scale evolution apparent in the graphs has to be made statistically significant and then further historical analysis will be necessary. Last, but not least, the comparison and cross-examination with the options market may provide important clues\citep{bates90}. 
 
 \section*{Acknowledgements}
 This work or its preliminary versions were presented in seminars in Paris, Rome and Trieste during the years 2009-2010 and in Buenos Aires in 2011. We gratefully acknowledge useful comments by J.P. Bouchaud, M. Potters, G. Biroli, J-P Nadal, M. Mezard, S. Franz, A. De Martino, L. Pietronero, G. Parisi, J. Doyne Farmer, F. Lillo, M. Marsilli, R. Perazzo, D. Heymann. I specially thank F. Lillo, R. Mantegna, E. Marinari and A. Tedeschi for the data. The ideas on Abduction developed slowly in long conversations with D. Amati.

\end{document}